\documentclass[aps,prl,groupedaddress,twocolumn,showpacs]{revtex4-1}
\usepackage{graphicx}
\usepackage{dcolumn}
\usepackage{latexsym}
\usepackage{amsmath, amsthm, amssymb}
\usepackage{epsfig}
\usepackage{latexsym}
\usepackage{bm}

\newcommand{\eq}[1]{Eq.~(\ref{#1})}
\newcommand{\fig}[1]{Fig.~\ref{#1}}

\begin{document}

\title{Jamming in a lattice model of stochastically interacting agents with a field of view}

\author{Shakti N. Menon, Trilochan Bagarti and Abhijit Chakraborty}
\affiliation{The Institute of Mathematical Sciences, CIT Campus, Taramani, Chennai 600113, India}

\date{\today}

\begin{abstract}
We study the collective dynamics of a lattice model of stochastically interacting agents with a weighted field of vision. We assume that agents preferentially interact with neighbours, depending on their relative location, through velocity alignments and the additional constraint of exclusion. Unlike in previous models of flocking, here the stochasticity arises intrinsically from the interactions between agents, and its strength is dependent on the local density of agents. We find that this system yields a first-order jamming transition as a consequence of these interactions, even at a very low density. Furthermore, the critical jamming density is found to strongly depend on the nature of the field of view.
\end{abstract}


\maketitle
%__________________________________________________________________________________________________%

\section{Introduction}
One of the most commonly observed manifestations of self-organized behaviour is collective motion, which occurs across length scales in a wide range of biological systems such as bacterial colonies, insect swarms, bird flocks, animal herds, etc~\cite{Vicsek2012}. In the last two decades, considerable effort has been made towards understanding the mechanism by which local rules that govern the dynamics of individual agents can impact global order~\cite{Cavagna2014,Romanczuk2012,Bhattacharya2016}. A pioneering work in this regard was the flocking model of self-propelled particles proposed by Vicsek~\emph{et al.}~\cite{Vicsek1995}, in which each particle attempts to align its direction to that of the local mean, in the presence of additive noise which accounts for the stochasticity in the alignments. On increasing the noise strength, this simple model exhibits a transition from an ordered to a disordered state~\cite{Gregoire2004,Baglietto2008}. However, the noise strength in this system is constant, spatially invariant, and is independent of the dynamics at the individual level. Most subsequent models of flocking have utilized a similar paradigm to incorporate the effects of noise on the collective dynamics. This rests on the underlying assumption that the uncertainty in the dynamics of an agent is independent of the density of its local environment, which may not necessarily be valid.

An important aspect of flocking that becomes increasing significant at higher densities is exclusion (i.e. hard-core repulsion)~\cite{Czirok1996,Romanczuk2009}. When this repulsive interaction is included, jamming can arise above a certain critical density~\cite{Bechinger2016}. The phenomenon of jamming has been studied extensively in contexts as diverse as road traffic~\cite{Helbing2001}, molecular motors on microtubules~\cite{Leduc2012} and crowd evacuation~\cite{Helbing2000}. In particular, numerous models of crowd dynamics have reported a jamming transition at a critical density~\cite{Piazza2010,Henkes2011,Jelic2012}. Lattice-gas models provide a convenient framework for incorporating exclusion and also exhibit a similar jamming transition~\cite{Dickman2001, Szolnoki2002, Braun2005, Toninelli2006, Potiguar2006}. Here, agents attempt random walks on a lattice and only interact via exclusion. These agents are not subject to the types of cooperative interactions that facilitate flocking, such as velocity alignment. Although flocking has explicitly been studied on a lattice~\cite{Csahok1995, Raymond2006}, to our knowledge exclusion has not been considered in models of this nature. In addition, we note that the phenomenologically distinct effects of cooperation and exclusion have not jointly been explored from a jamming perspective. Furthermore, a common assumption in such models is that the agents have a visibility that is radially symmetric and unweighted, i.e. a spatially uniform field of view. Although some off-lattice models have considered a spatially variable field of view for the agents~\cite{Couzin2002,Hemelrijk2012,Romensky2014,Pearce2014}, it is of potential interest to explore how the jamming transition of agents on a lattice may depend on such a constraint.

In this Letter, we introduce a lattice model that describes the collective behaviour of agents with a field of view in the presence of exclusion and alignment interactions. Here, the stochasticity in the dynamics arises purely from uncertainties present at the individual level, rather than from additive noise, i.e. the dynamics is intrinsically stochastic. Although additive noise can account for environmental fluctuations, as in the case of Brownian particles in a thermal environment \cite{vanKampen1992}, a more significant contribution to the fluctuations in the case of active particles originates from the interactions between particles. For instance, in flocking dynamics one expects that the random motion of an agent depends on the density of its local neighbourhood. The number of interactions by a given agent is proportional to the number of agents in its vicinity. Hence, the strength of randomness varies with the instantaneous density of the local neighbourhood, unlike additive noise which is independent of the local density. A principle that could account for this type of randomness has not been formulated thus far for the case of flocking dynamics of self-propelled particles. Furthermore, since the stochasticity is generated due to the uncertainty in decision-making at the level of individual agents, the noise term and the deterministic term may not be separable. Hence, to account for uncertainties in the interactions between agents we propose a velocity update rule that is intrinsically stochastic.

We find that our model gives rise to various jamming transitions even at very low densities. Here, we have investigated two cases, each corresponding to a different configuration of the field of view for all agents. It is observed that the field of view is crucial in determining the nature of transitions and the critical jamming density. Despite being a simple model, we observe that it yields a rich variety of spatiotemporal phenomena.

%====================================================================================================%
\section{Model}
We consider a set of $N$ agents on a square lattice of size $L \times L$. At every time step, each agent attempts to hop to the neighbouring site that is in the direction of its hopping velocity. The state of the system at time $t$ is described by $(\mathbf{x}_1(t), \mathbf{x}_2(t),\ldots,\mathbf{x}_N(t),\mathbf{v}_1(t), \mathbf{v}_2(t),\ldots,\mathbf{v}_N(t))$, where $\mathbf{x}_i(t)$ denotes the position of agent $i$ and $\mathbf{v}_i(t)$ is its hopping velocity. The agents have discrete velocities $\mathbf{v}_i=(\eta_x, \eta_y)^T$, where $\eta_{x(y)}=0,\pm 1$ and $\mathbf{v}_{i}\neq(0,0)^T$. The dynamics of the system constitutes agents hopping to a nearest or next nearest neighbour site, subject to exclusion, and velocity-dependent interactions (see the schematics in the Supplementary Information for an overview of the velocity alignment update rule). The rules governing the hopping of agent $i$ at step $t+1$ are
%____________________________________________%
\begin{subequations}
\begin{align} 
\label{vel-update}
\mathbf{v}_i(t+1) &= \mathbf{v}' \in \Omega_i \text{~~with probability~~} P(\mathbf{v}'|\mathbf{v}_i(t)),\\
\label{pt-update}
\mathbf{x}_i(t+1) &= \mathbf{x}_i(t) + \mathbf{v}_i(t+1)\,n(\mathbf{x}_i(t)+\mathbf{v}_i(t+1)),
\end{align}
\label{eq-update}
\end{subequations}
%____________________________________________%
where $\Omega_i$ is the set of velocities in the Moore neighbourhood of $\mathbf{x}_i(t)$,
$P(\mathbf{v}'|\mathbf{v}_i(t))$ is the transition probability for the update of velocity from  $\mathbf{v}_i(t)$ to $\mathbf{v}'$, and $n(\mathbf{x})$ incorporates the exclusion, i.e. $n(\mathbf{x})=0$ if the site $\mathbf{x}$ is occupied and 1 otherwise. The initial condition is $\mathbf{x}_i(0)=\mathbf{x}_i^{0}$ and $\mathbf{v}_i(0)=\mathbf{v}_i^{0}$ $\forall i=1\ldots N$. The set $\Omega_i$ in \eq{vel-update} contains the velocities assigned to the eight neighbouring sites of $\mathbf{x}_{i}$. If a site is unoccupied, that site is assigned the velocity of the occupying agent. If a site is empty, the velocity assigned to that site in $\Omega_{i}$ is either the velocity of the agent at $\mathbf{x}_i(t)$ with probability $p_{\text{self}}$, or a random velocity with probability $1-p_{\text{self}}$. This ensures that an agent $i$ with velocity $\mathbf{v}_i(t)$ that has no neighbour will, at step $t+1$, retain its velocity with probability $p_{\text{self}}$ and be assigned a random velocity with probability $1-p_{\text{self}}$. Thus, for a single isolated agent, \eq{vel-update} describes a random walk with nearest and next-nearest hopping for $p_{\rm self}=0$, and is ballistic motion for $p_{\rm self}=1$, with $\mathbf{x}_{i}(t+1)=\mathbf{x}_{i}(0)+\mathbf{v}_{i}(0)\,t$, $\mathbf{v}_{i}(t+1)=\mathbf{v}_{i}(0)$. For intermediate values of $p_{\rm self}$ the dynamics constitutes a combination of deterministic and random components. We note that \eq{eq-update} is non-Markovian, which is an necessary feature of collective behaviour that arises from the proposed update rule.

%------------------------------------------------------------------------------------------------%
\begin{figure}
\includegraphics{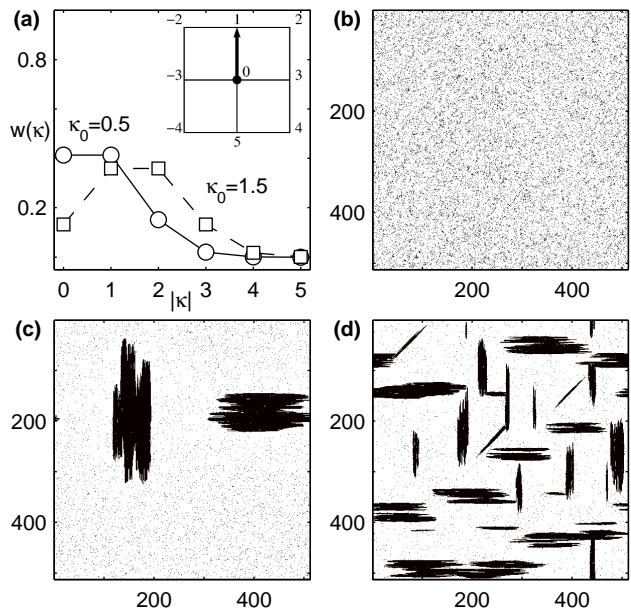}
\caption{(a) Gaussian weights $w(\kappa)=\exp(-(|\kappa|-\kappa_0)^2/2\sigma^2)$ for the transition probabilities used in the model, displayed for the cases $p_{\rm self}=0.75$, $\sigma=1$, $\kappa_{0}=0.5$ (circles) and $\kappa_{0}=1.5$ (squares). [inset] Schematic that displays the locations of the relative neighbours $\kappa$ for a agent at $0$ that is pointing in the direction of $1$.
(b-d) Snapshots of a system of agents on a $512\times512$ lattice for the case $p_{\rm self}=0.75$ and $\kappa_{0}=0.5$ at time $t=10^4$ for the densities (b) $\rho=0.1$, (c) $\rho=0.12$, and (d) $\rho=0.2$.
}
\label{fig1}
\end{figure}
%------------------------------------------------------------------------------------------------%

The transition probability $P(\mathbf{v}'|\mathbf{v}_i(t))$ in \eq{vel-update} is symmetric about the direction of $\mathbf{v}_i$ (c.f. \fig{fig1}a[inset], where the origin corresponds to $\mathbf{x}_i$). This function only depends on the relative position (with respect to the site $\mathbf{x}_i$) of the agent with velocity $\mathbf{v}'$. Note that the transition probability for an agent is spatially independent, i.e. it does not explicitly depend on its position on the lattice. Furthermore, the transition probability $P(\mathbf{v}'|\mathbf{v}_i)=P(\mathbf{v}''|\mathbf{v}_i)$, for all choices of $\mathbf{v}'$ and $\mathbf{v}''$ for a given neighbor. The transition probability can be written in terms of weights $w(\kappa)$, such that $P(\mathbf{v}_{\kappa}|\mathbf{v}_i) = w(\kappa)/\sum_{\kappa'} w(\kappa')$, where $\mathbf{v}_{\kappa}$ is the velocity at the neighbouring site $\kappa$. Here we use a shifted Gaussian function for the weights: $w(\kappa)=\exp(-(|\kappa|-\kappa_0)^2/2\sigma^2)$, where $\kappa=0,\pm1\ldots,\pm 5$ denotes the index of the neighbouring site relative to $\mathbf{x}_i$, the shift $\kappa_0$ denotes the relative preference in velocity alignment and $\sigma$ is the width of the field of view. The indices of the neighbors are shown in \fig{fig1}(a)[inset] for the case where $\mathbf{v}_i=(0,1)^T$ points to the north, and for other cases the indices $\kappa$ may be numbered accordingly.

The values of the shift $\kappa_0$ and the width $\sigma$ determines the properties of the dynamics. When $\sigma \rightarrow \infty$ the transition probability $P(\mathbf{v}_{\kappa}|\mathbf{v}_i)=1/9$ for all $\mathbf{v}_{\kappa}$. That is, the agent aligns with any of its neighbours with equal probability. Similarly, when $\sigma \rightarrow 0$ and $\kappa_0=0$, $P(\mathbf{v}_{\kappa}|\mathbf{v}_i)=1$ for $\kappa=0$, and is $0$ otherwise. This implies that the agent does not align its velocity with any of its neighbours. In the simulations that follow, we choose the values $\kappa_0=0.5$ and $\kappa_0=1.5$ for different values of $\sigma$. For the case $\kappa_0=0.5$ we note from \fig{fig1}(a), that the weights $w(0)=w(\pm 1)\geq w(\kappa)$ for all $|\kappa|>1$. However, for $\kappa_0=1.5$, the weights $w(\pm 1)=w(\pm 2) \geq w(\kappa)$ for all $|\kappa|>2$ and $\kappa=0$. We note that $\sigma$ effectively corresponds to the nature of the field of view of the agent with weights specified by $w(\kappa)$. Furthermore, given an agent that points in a direction $\mathbf{v}$, $\kappa_{0}$ determines the relative preference given to its neighbouring sites in a frame of reference aligned with $\mathbf{v}$.

{\it Numerical simulations:}
At time $t=0$, $\mathbf{x}_i$ is uniformly distributed on the lattice and $\mathbf{v}_i$ is assigned one of the eight possible velocities with equal probability $\forall i=1\ldots N$. At time step $t$, the velocity of each agent is first updated synchronously by rule \eq{vel-update}. The positions are then updated by rule \eq{pt-update} in a random sequence in order to avoid correlations, since the updated velocity $\mathbf{v}_i(t+1)$ appears on the right hand side of \eq{pt-update}. Periodic boundary conditions are used along both directions. In our simulations, we have used lattices of size $L=64, 90, 128, 180,\text{and~}256$. The values of the shift  used here are $\kappa_{0}=0.5$ and $\kappa_{0}=1.5$ with $\sigma=0.1,1,10$, and the probability is set to $p_{\rm self}=0.75$ unless specified otherwise.

%====================================================================================================%
\section{Results}

We find that agents begin to cluster as we increase the density $\rho(=N/L^2)$, and jamming can abruptly emerge for certain $\sigma$, even for very low density. In \fig{fig1}(b-d), we show snapshots from a simulation for values of $\rho$ around this jamming transition. We observe that large clusters instantaneously emerge above a critical density $\rho_{c}$. Unlike in active particle systems, where clusters can arise due to the presence of attractive forces, here clusters form due to exclusion and velocity alignment. At low $p_{\rm self}$ the clusters are less elongated than that shown in \fig{fig1}, due to the fact that in this case the velocities in the unoccupied sites are more likely to be random (see Supplementary Information).

In order to quantify the collective motion of the system, we define the mobility of the population $\mu(t)$ at a given time $t$ to be $\mu(t) := N^{-1} \sum_i (1-\delta_{\mathbf{x}_i(t),\,\mathbf{x}_i(t+1)})$. We set our order parameter to be the asymptotic mean mobility, which is defined as $\langle \mu \rangle := \langle \mu(t) \rangle$, $t \rightarrow \infty$, where $\langle \cdot \rangle$ denotes average over an ensemble. To investigate the jamming transition, we consider two cases that correspond to different fields of view of the agents, that is $\kappa_{0}=0.5$ and $\kappa_{0}=1.5$ (c.f. \fig{fig1}(a)). For $\kappa_{0}=0.5$ each agent picks the site that is directly in front of it (i.e. in the direction in which it is pointing) with equal probability to picking its own site, whereas for $\kappa_{0}=1.5$ all three sites in front are picked with an equal probability that is higher than that for picking its own site.

%------------------------------------------------------------------------------------------------%
\begin{figure}
\begin{tabular}{cc}
\includegraphics{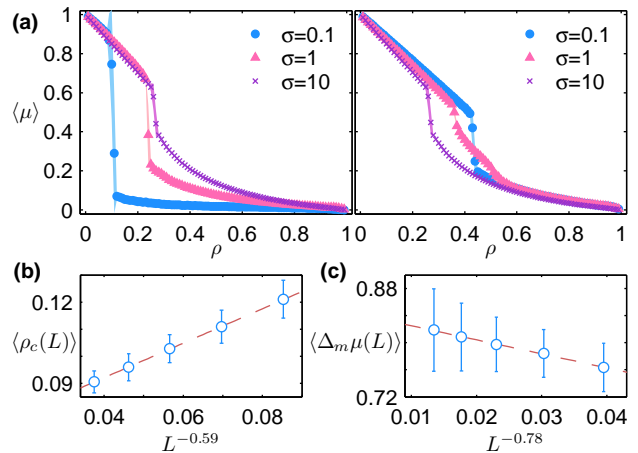}
\end{tabular}
\caption{Simulations of agents on a $128\times 128$ lattice for the case $p_{\rm self}=0.75$. (a) Mobility $\mu$ as a function of density $\rho$, averaged over 100 trials, for values of $\kappa_{0}$ corresponding to two different transition probabilities, namely [left] $\kappa_{0}=0.5$ and [right] $\kappa_{0}=1.5$. Results are shown for the cases $\sigma=0.1$ (filled circles), $\sigma=1$ (triangles) and $\sigma=10$ (crosses). The shaded regions indicate the extent of fluctuation of the standard deviation. Note that for the case $\kappa_{0}=0.5$  the jamming transition point (i.e. the critical density $\rho_c$) shifts towards lower values on decreasing $\sigma$, whereas the reverse trend is seen for $\kappa_{0}=1.5$.
(b) Critical density $\rho_{c}$ as a function of system size $L$, averaged over $1000$ trials for the case $p_{\rm self}=0.75$, $\kappa_0=0.5$, $\sigma=0.1$, and shown as empty circles. The bars indicate the standard deviation and the least-squared fit of the data (shown as a dashed line) follows $0.067+0.636\,L^{-0.59}$.
(c) Maximum jump size $\Delta_{m}\mu$ in the order parameter $\mu$ as a function of system size $L$, averaged over $1000$ trials for the same data used in in (b), and shown as empty circles. The bars indicate the standard deviation and the least-squared fit of the data (shown as a dashed line) follows $0.846-2.073\,L^{-0.78}$.
}
\label{fig2}
\end{figure}
%------------------------------------------------------------------------------------------------%

The order parameter $\langle \mu \rangle$ is shown for different values of $\sigma$ for $p_{\rm self}=0.75$ in \fig{fig2}(a). For $\kappa_{0}=0.5$ (\fig{fig2}(a)[left]), we observe that the transition point $\rho_{c}$ shifts to the right as we increase $\sigma$. However, the converse is observed for $\kappa_{0}=1.5$ (\fig{fig2}(a)[right]). Furthermore, we observe that for $\sigma=10$, the order parameter $\langle \mu \rangle$ is identical for $\kappa_{0}=0.5$ and $1.5$, as it is sufficiently close to that of the limiting case $\sigma\rightarrow\infty$, where the field of view extends equally in all directions. For low $p_{\rm self}$, similar behaviour is observed for $\kappa_{0}=0.5$, but not for $\kappa_{0}=1.5$ (see Supplementary Information). This implies that $p_{\rm self}$ crucially determines the nature of the shift in $\rho_{c}$. The behaviour of $\langle \mu \rangle$ in all the transitions shown in \fig{fig2}(a) has been found to decrease monotonically with $\rho$ and it is approximately linear, i.e. $\langle\mu\rangle\sim 1-a\rho$, for $\rho<\rho_c$ and $\langle\mu\rangle\sim (1-a \rho)\,\exp(-b\rho)$ for $\rho>\rho_c$, where $a$ and $b$ are constants. On increasing the system size $L$, we find that the critical density $\rho_{c}$ scales as $L^{-0.59}$, and in the asymptotic limit $L\rightarrow\infty$, $\rho_{c}(L)\approx0.067$ (see \fig{fig2}(b)). It is surprising that even at this low critical density, the system shows a sharp jamming transition. We note that this is even lower than the value obtained for a lattice-gas system with nearest-neighbour exclusion, namely $\rho_c\simeq 0.263$~\cite{Dickman2001}. Furthermore, we observe that the average value of the maximum jump in the order parameter $\Delta_{m}\mu$, scales as $L^{-0.78}$ and in the asymptotic limit $L\rightarrow\infty$, $\Delta_{m}\mu(L)\approx0.846$  (see \fig{fig2}(c)). This is indicative of a first-order transition -- a claim that we validate in the following analysis.

A finite size scaling of the order parameter $\langle\mu\rangle$ is shown in \fig{fig3}(a) for the sharp transition corresponding to the case $\kappa_{0}=0.5$ and $\sigma=0.1$. We find that data collapse is observed by scaling $\rho-\rho_{c}(L)$ by a factor of $L^{\alpha}$ with $\alpha=0.5$. To characterize the nature of this transition we compute the Binder cumulant $G=1 - \langle \mu^4\rangle/3\langle \mu^2\rangle^2$. As shown in the inset of \fig{fig3}(a)), the binder cumulant is negative at $\rho_{c}(L)$ for all system sizes considered, which in conjunction with \fig{fig2}(c) substantiates our claim of a first-order transition. Although we only find a negative binder cumulant for the case $\kappa_{0}=0.5$ and $\sigma=0.1$, we observe that the distribution of order parameters for all the six transitions displayed in \fig{fig2}(a) are bimodal (see Supplementary Information), which suggests that all the transitions observed for the parameters considered here are first order.

We observe that the emergence of clusters is characterized by a sharp drop in $\langle\mu\rangle$ after an initial transient period of of higher mobility. We define by $p(\tau_d)$ the probability density for the first drop in mobility below a chosen threshold ($\mu_{\rm th}=0.5$) at time $\tau_{d}$. In the vicinity of $\rho_{c}$, the probability distribution is found to be exponentially decaying (see \fig{fig3}(b)), with the rate of decay larger at higher $\rho$. To understand the dynamics of individual agents at $\rho_{c}$, we investigate the diffusive properties of randomly chosen tracer particles. As seen in \fig{fig3}(c) the probability density of displacement $\Delta x$ of a tracer particle is approximately Gaussian at small $t$ and is exponentially decaying for larger $t$. The latter case obeys the exponential distribution $p(\Delta x)\sim \exp(-|\Delta x|/t^{\beta})$, where $\beta=0.21$ since the mean-square displacement is found to be $\langle |\Delta \mathbf{x}|^2 \rangle\sim t^{0.43}$ (see \fig{fig3}(d)). This confirms that the tracers show a sub-diffusive behaviour at larger time. As the first drop probability density $p(\tau_d)$ shows an exponential decay, there is a higher probability of formation of jammed clusters at larger $t$, which appears to give rise to this sub-diffusive behaviour.

%------------------------------------------------------------------------------------------------%
\begin{figure}
\begin{tabular}{cc}
\includegraphics{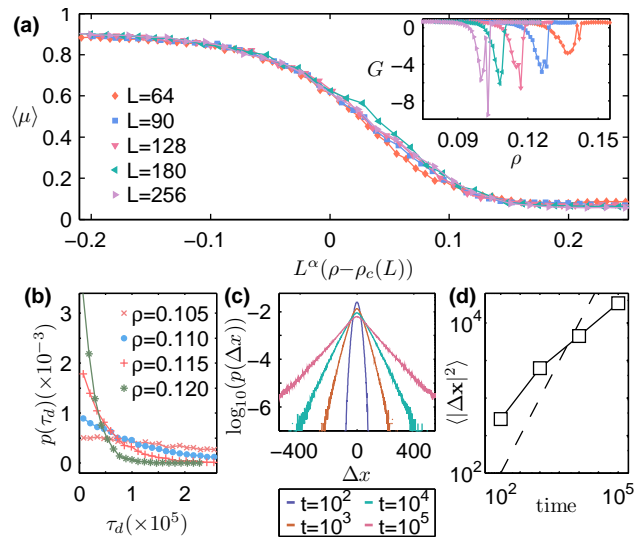}
\end{tabular}
\caption{Detailed statistics of the collective dynamics for the case $p_{\rm self}=0.75$, $\kappa_0=0.5$, $\sigma=0.1$.
(a) Finite size-scaling of the mobility $\langle\mu\rangle$, averaged over $1000$ trials, shown as a function of the density $\rho$ for different lattice sizes $L$ with exponent $\alpha=0.5$. [inset] Corresponding Binder cumulants $G$ for the curves shown in (a).
(b) Probability distributions of first-drop times $\tau_{d}$ shown over a range of values of $\rho$,  calculated over $10^4$ trials on a $128\times 128$ lattice in each case.
(c) Probability distributions of displacement along the $x$-direction, $\Delta x$, at different times $t$, for $10^{3}$ tracer agents on a $200\times 200$ lattice with $\rho=0.125$, calculated over $10^{4}$ trials.
(d) Mean squared displacement $\langle |\Delta \mathbf{x}|^2 \rangle$  as a function of time for the same data used in (c), shown as empty squares. The dashed line which represents the mean squared displacement for normal diffusion is shown for reference.
}
\label{fig3}
\end{figure}
%------------------------------------------------------------------------------------------------%

%====================================================================================================%
\section{Conclusions}

In this work we have investigated the collective behavior that arises through the interplay of exclusion and velocity alignment of agents on a lattice.
The model incorporates randomness in the dynamics by assuming stochastic alignment interactions between agents with a variable field of view. Thus, unlike in previous studies of collective motion, here the stochasticity has been considered to arise naturally from the alignment interactions between the agents. This captures the intrinsic nature of stochasticity in such systems, and we believe that it should be an essential ingredient when formulating a flocking model. We have shown that a system governed by these rules exhibits a first-order jamming transition in the mean mobility. The nature of the transitions depend crucially on the relative preference in velocity alignments, the width of the field of view $\sigma$, and the probability $p_{\rm self}$ that an agent that chooses a vacant site during the alignment process will retain its velocity rather than be assigned a random velocity. We observe that for a large value of $p_{\rm self}$ the transition for a fixed value of $\sigma$ will occur at a lower density if less relative preference is given to a neighbour ($\kappa_0=0.5$), or a larger density if the neighbour is given a higher preference ($\kappa_0=1.5$). Thus, depending on the relative preference, cooperative behaviour can either increase or decrease the efficiency of herd movement. We infer that the balance between exclusion and cooperativity is crucially significant in determining the nature of collective motion. These results may find applications in diverse areas, including studies of crowd control and of quadruped movement. 

While we have demonstrated that this system exhibits a first-order transition for a narrow field of view, it appears that this model will exhibit a sharp transition regardless of the system parameters. This, along with the effect of $p_{\rm self}$ on the nature of the transition, shall be investigated in a future work. Furthermore, the framework employed here will be extended to study flocking behaviour in off-lattice models.

%====================================================================================================%

\acknowledgements
We would like to thank V. Sasidevan and Gautam Menon for helpful discussions. SNM is supported by the IMSc Complex Systems Project ($12^{\rm th}$ Plan). The simulations and computations required for this work were supported by the Institute of Mathematical Science's High Performance Computing facility (hpc.imsc.res.in) [nandadevi and satpura], which is partially funded by DST.

%____________________________________________________________________________________________________%

%____________________________________________________________________________________________________%

\pagebreak
\begin{table*}
{\large {\bf SUPPLEMENTARY INFORMATION}}
\end{table*}
\setcounter{figure}{0}
\renewcommand\thefigure{S\arabic{figure}}  
\renewcommand\thetable{S\arabic{table}}

%------------------------------------------------------------------------------------------------%
\begin{figure*}
\begin{tabular}{cc}
\centering{\includegraphics[width=0.49\textwidth]{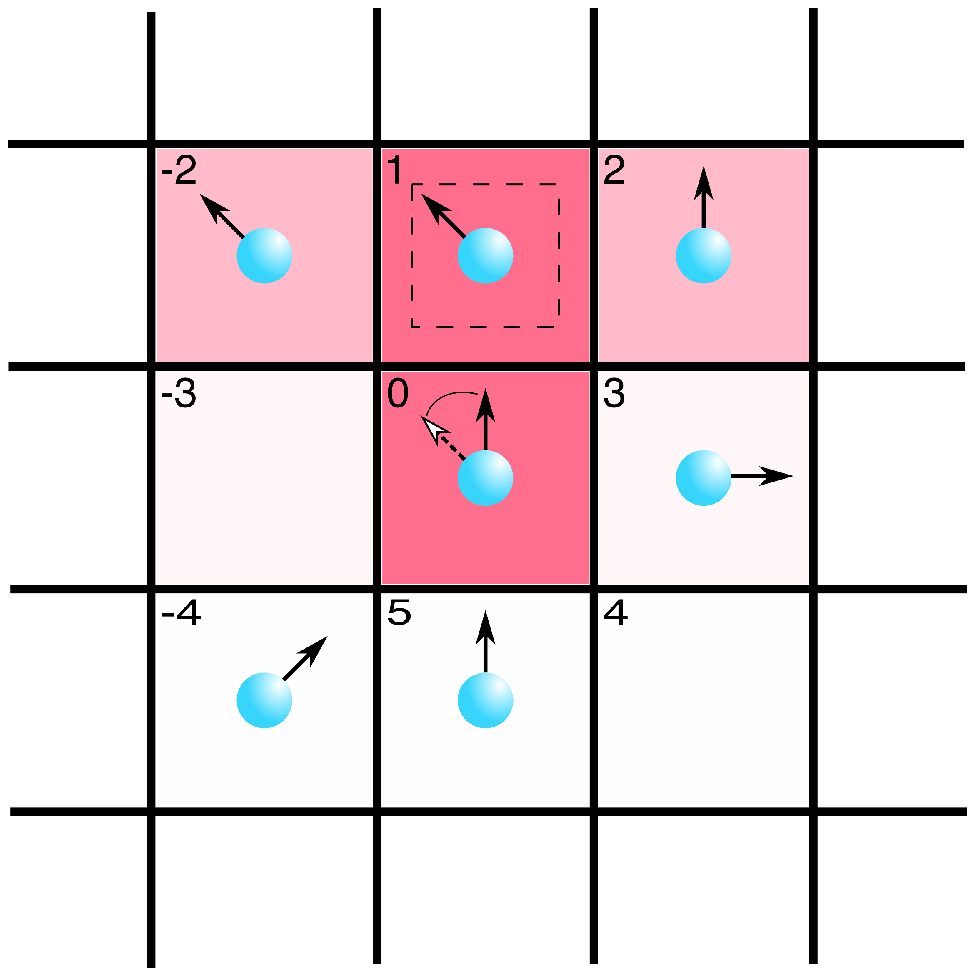}}&
\centering{\includegraphics[width=0.49\textwidth]{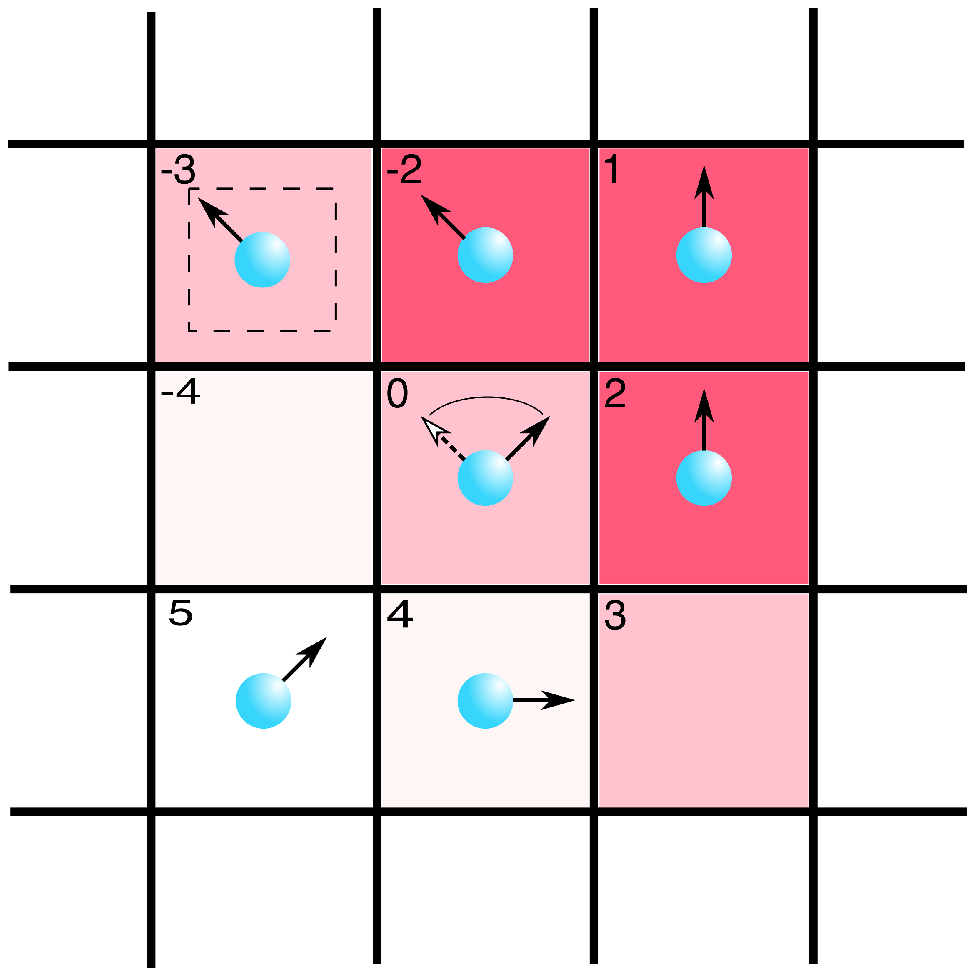}}
\end{tabular}
\caption{Schematics illustrating the stochastic alignment update rule used in the model. The two
panels correspond to the field of view for a given agent for the cases [left] $\kappa_{0}=0.5$,
and [right] $\kappa_{0}=1.5$. In each case, an agent is located at the centre of a $3\times 3$
grid, and other agents occupy the nearest and next-nearest neighbouring sites
of the central agent. Each agent is represented by a coloured ball, and its direction is indicated
by a black arrow. In the left panel the central agent points to the north and in the right it points north-west. The values of $\kappa$ in the neighbouring sites are chosen accordingly. The lattice sites are coloured with an intensity that is proportional to the
weight of the probability distribution $w(\kappa)=\exp(-(|\kappa|-\kappa_0)^2/2\sigma^2)$ where,
for the schematics shown here, we set $\sigma=1$ and $\kappa=0,\pm 1\ldots \pm 5$. The locations
of the sites corresponding to each $\kappa$ are chosen relative to the direction of the central
agent (assumed to be at $\kappa=0$). The value of $\kappa$ corresponding to each site in the two
panels is indicated in the top left corner of each site. The central agent is more likely to
select a neighbouring site with a more intense colour. The dotted box indicates the site randomly
chosen by the central agent. On making this choice, the central agent aligns its velocity to that
of any agent that may be present at that site. The updated direction of the agent is indicated by
a dotted arrow. If the chosen site is empty, the agent retains its original velocity with
probability $p_{\rm self}$ and is assigned a random velocity with probability $1-p_{\rm self}$.
}
\label{fig1}
\end{figure*}
%------------------------------------------------------------------------------------------------%

%------------------------------------------------------------------------------------------------%
\begin{figure*}
\centering{\includegraphics{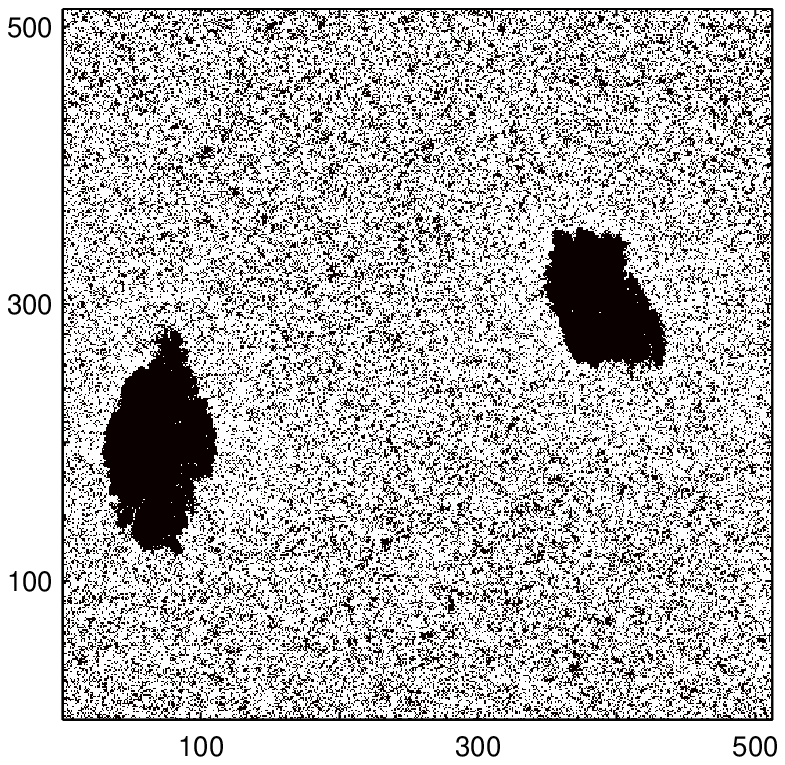}}
\centering{\includegraphics{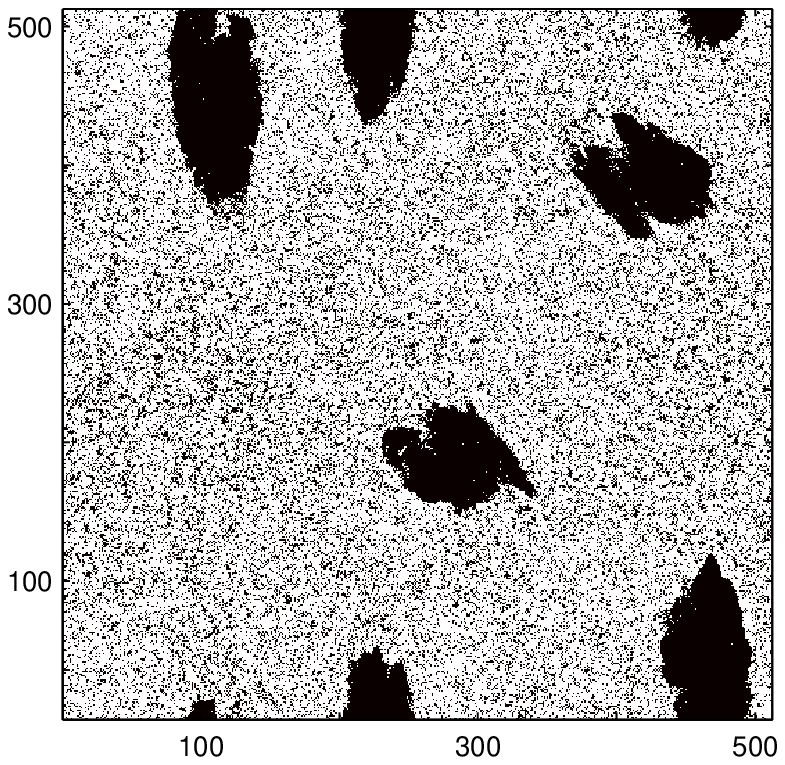}}
\caption{
Representative jammed clusters obtained for different choices of the parameter $p_{\rm self}$.
Snapshots of a system of agents on a $512\times512$ lattice with $\sigma=1$ and density
$\rho=0.3$ are displayed for $\kappa_{0}=0.5$ at time $t=2\times10^4$ for the cases [left]
$p_{\rm self}=0.1$ and [right] $p_{\rm self}=0.25$. It can be observed that the jammed clusters
are less elongated for smaller values of $p_{\rm self}$.
}
\label{fig2}
\end{figure*}
%------------------------------------------------------------------------------------------------%

%------------------------------------------------------------------------------------------------%
\begin{figure*}
\centering{\includegraphics[width=0.9\textwidth]{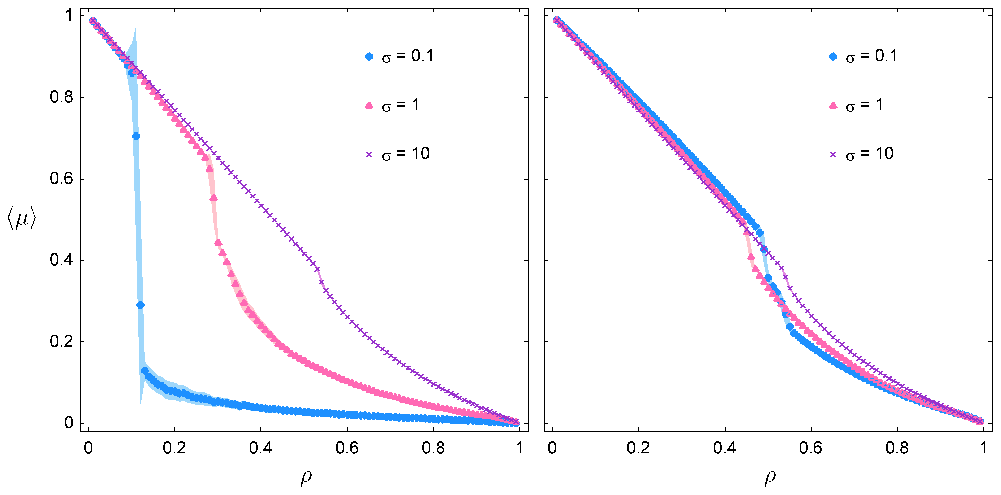}}
\caption{
Phase transitions in the average mobility of agents on a $128\times 128$ lattice for the case 
$p_{\rm self}=0.25$. The mobility $\mu$, averaged over 100 trials, is displayed as a function of
density $\rho$ for values of $\kappa_{0}$ corresponding to two different transition probabilities,
namely [left] $\kappa_{0}=0.5$ and [right] $\kappa_{0}=1.5$. Results are shown for the cases
$\sigma=0.1$ (filled circles), $\sigma=1$ (triangles) and $\sigma=10$ (crosses). The shaded
regions indicate the extent of fluctuation of the standard deviation. It can be seen that these
results are qualitatively similar to those obtained for the case $p_{\rm self}=0.75$ (displayed in
the main text).
}
\label{fig3}
\end{figure*}
%------------------------------------------------------------------------------------------------%

%------------------------------------------------------------------------------------------------%
\begin{figure*}
\centering{\includegraphics{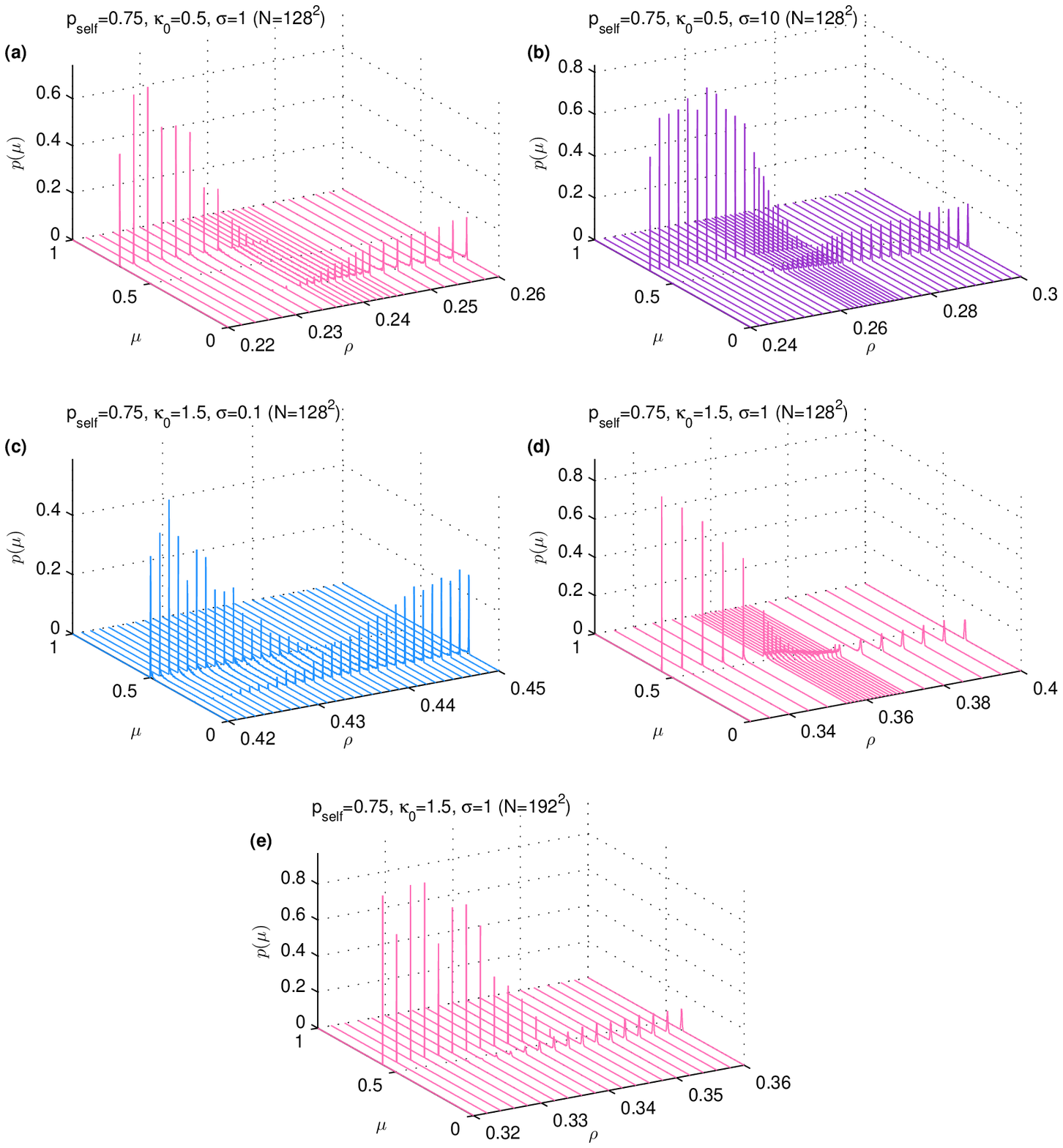}}
\caption{Order parameter distributions across the jamming transitions for the case
$p_{\rm self}=0.75$ displayed in Fig.~2(a) of the main text. In each panel, we display the
probability $p(\mu)$ (calculated over $1000$ trials) that the value of the order parameter $\mu$
is a certain value between $0$ and $1$, for a range of values of $\rho$ around the corresponding
jamming transitions. The first four cases shown are for a $128\times 128$: (a) $\kappa_{0}=0.5$,
$\sigma=1$, (b) $\kappa_{0}=0.5$, $\sigma=10$, (c) $\kappa_{0}=1.5$, $\sigma=0.1$ and
(d) $\kappa_{0}=1.5$, $\sigma=1$. In cases (a)-(c), we see a bimodal distribution near the
transition point, indicative of a first-order transition. Although the transition in (d) appears
to be continuous, we observe that this is a system size effect. We observe in (e) that for a
larger system size ($192\times 192$) we once again obtain a bimodal distribution near the
transition. For ease of comparison with the main text, the distributions in each panel are
coloured according to the corresponding curves in Fig.~2(a) of the main text.
}
\label{fig4}
\end{figure*}
%------------------------------------------------------------------------------------------------%

%====================================================================================================%


\begin{thebibliography}{00}
\bibitem{Vicsek2012} T.~Vicsek and A.~Zafeiris, {\it Phys. Rep.}, {\bf 517} (2012) 71.
\bibitem{Cavagna2014} A.~Cavagna and I.~Giardina, {\it Annu. Rev. Condens. Matter Phys.}, {\bf 5} (2014) 183.
\bibitem{Romanczuk2012} P.~Romanczuk~\emph{et al.}, {\it Eur. Phys. J.-Spec. Top.}, {\bf 202} (2012) 1.
\bibitem{Bhattacharya2016} K.~Bhattacharya and A.~Chakraborty, arXiv preprint arXiv:1504.02022 (2016).
\bibitem{Vicsek1995} T.~Vicsek \emph{et al.}, {\it Phys. Rev. Lett.}, {\bf 75} (1995) 1226.
\bibitem{Gregoire2004} G.~Gr\'{e}goire and H.~Chat\'{e}, {\it Phys. Rev. Lett.} {\bf 92} (2004) 025702.
\bibitem{Baglietto2008} G.~Baglietto and E.~Albano, {\it Phys. Rev. E}, {\bf 78} (2008) 021125.
\bibitem{Czirok1996} A.~Czir\'{o}k \emph{et al.}, {\it Phys. Rev. E}, {\bf 54} (1996) 1791.
\bibitem{Romanczuk2009} P.~Romanczuk, I.~D.~Couzin and L.~Schimansky-Geier, {\it Phys. Rev. Lett.}, {\bf 102}  (2009) 010602.
\bibitem{Bechinger2016} C.~Bechinger \emph{et al.}, arXiv preprint arXiv:1602.00081 (2016).
\bibitem{Helbing2001} D.~Helbing, {\it Rev. Mod. Phys.}, {\bf 73} (2001) 1067.
\bibitem{Leduc2012} C.~Leduc {\emph et al.}, {\it Proc. Nat. Acad. Sci. USA}, {\bf 109} (2012) 6100.
\bibitem{Helbing2000} D.~Helbing, I.~Farkas and T.~Vicsek, {\it Nature}, {\bf 407} (2000) 487.
\bibitem{Piazza2010} F.~Piazza, {\it Phys. Rev. E}, {\bf 82} (2010) 026111.
\bibitem{Jelic2012} A.~Jelic \emph{et al.}, {\it Phys. Rev. E}, {\bf 85} (2012) 036111.
\bibitem{Henkes2011} S.~Henkes, Y.~Fily, and M.~C.~Marchetti, {\it Phys. Rev. E}, {\bf 84} (2011) 040301(R).
\bibitem{Braun2005} O.~M.~Braun and B.~Hu, {\it Phys. Rev. E}, {\bf 71} (2005) 031111.
\bibitem{Potiguar2006} F.~Q.~Potiguar and R.~Dickman, {\it Eur. Phys. J. B}, {\bf 52} (2006) 83.
\bibitem{Szolnoki2002} A.~Szolnoki and G.~Szabo, {\it Phys. Rev. E}, {\bf 65} (2002) 047101.
\bibitem{Toninelli2006} C.~Toninelli, G.~Biroli and D. S. Fisher, {\it Phys. Rev. Lett.}, {\bf 96} (2006) 035702.
\bibitem{Dickman2001} R.~Dickman, {\it Phys. Rev. E}, {\bf 64} (2001) 016124.
\bibitem{Csahok1995} Z.~Csahok and T.~Vicsek, {\it Phys. Rev. E}, {\bf 52} (1995) 5297.
\bibitem{Raymond2006} J.~R.~Raymond and M.~R.~Evans, {\it Phys. Rev. E}, {\bf 3} (2006) 036112.
\bibitem{vanKampen1992} N.~G.~Van Kampen, {\it Stochastic processes in physics and chemistry} (Elsevier, 1992).
\bibitem{Couzin2002} I.~D.~Couzin \emph{et al.}, {\it J. Theor. Biol.} {\bf 218} (2002) 1.
\bibitem{Hemelrijk2012} C.~K.~Hemelrijk and H.~Hildenbrandt, {\it Interface Focus}, {\bf 2} (2012) 726.
\bibitem{Romensky2014} M.~Romensky, V.~Lobaskin and T.~Ihle, {\it Phys. Rev. E}, {\bf 90} (2014) 063315.
\bibitem{Pearce2014}D.~J.~Pearce \emph{et al.}, {\it Proc. Natl. Acad. Sci. USA}, {\bf 111} (2014) 10422.
\end{thebibliography}
\end{document}